\documentclass[twocolumn,prl,showpacs]{revtex4}
\usepackage{epsf}
\usepackage{color}
\usepackage{verbatim}
\usepackage{hyperref}
\usepackage{graphics}
\usepackage{graphicx}
\usepackage{natbib}
\usepackage{dcolumn}
\usepackage{bm}

\begin{document}
\title{Intimate relations between electronic nematic, d-density wave, and
d-wave superconducting states }
%\title{ The distinction between d-density wave and electronic nematic orders is lost?}
\author{Hae-Young Kee}
%\email{hykee@physics.utoronto.ca}
\affiliation{Department of Physics, University of Toronto, Toronto,
Ontario M5S 1A7 Canada}
\affiliation{School of Physics, Korea Institute for Advanced Study, Seoul 130-722, Korea }
\author{Hyeonjin Doh}
%\email{hdoh@physics.utoronto.ca}
\affiliation{Department of Physics, University of Toronto, Toronto,
Ontario M5S 1A7 Canada}
\author{Thomas Grzesiak}
\affiliation{Department of Physics, University of Toronto, Toronto,
Ontario M5S 1A7 Canada}
\date{\today}
\begin{abstract}
This paper consists of two important theoretical observations
on the interplay between $l=2$ condensates;  d-density wave (ddw), 
electronic nematic, and  d-wave superconducting states.
(1) There is SO(4) invariance at a transition between the
nematic and d-wave superconducting states. The nematic and d-wave pairing operators
can be rotated into each other by pseudospin SU(2) 
generators, which are s-wave pairing and electron density operators.
The difference between the current work and the previous O(4) symmetry at a transition between the ddw and d-wave
superconducting states\cite{nayak00PRB} is presented.
(2) The nematic and  ddw  operators transform into each other
under a unitary transformation.  Thus when a Hamiltonian is invariant under 
such a transformation, two states are exactly degenerate. 
The competition between the nematic and ddw states in the presence of a degeneracy-breaking term 
is discussed.
\end{abstract}
\maketitle

{\it Introduction} ---
Motivated by the discovery of exotic ordered states in strongly correlated materials,
the interplay between different order parameters has been of great interest.
In particular, competition and/or cooperation of the d-wave superconducting state 
and other nearby ordered states in high temperature cuprates
have been subjects of intensive theoretical research activities.
A few examples of nearby ordered states proposed in high temperature cuprates 
include the Neel antiferromagnet\cite{demler04RMP,chubukov03},
the ddw state(also called the staggered flux phase) 
\cite{chakravarty02PRB,affleck88PRB, palee07},
and the electronic nematic phase
\cite{kee03PRB,yamase00JPSJ, oganesyan01PRB, metzner00PRL, kivelson03RMP}.

Among these, the antiferromagnetic state is an s-wave particle-hole condensate, while
the ddw, electronic nematic, and d-wave superconducting states share a  common
d-wave feature; the d-wave superconducting state is formed by a condensation
of particle-particle pairs of $l=2$ relative angular momentum, while the ddw and
electronic nematic states are formed by condensations of particle-hole pairs of
$l=2$ relative angular momentum, but at different wavevectors.  

In this paper, we study whether there are 
intimate relations between  these $l=2$ condensates, which will shed 
light on our understanding of the competition and/or cooperation between them.
The d-wave superconducting, ddw, and electronic nematic states are represented by
their order parameters which capture the characteristic broken symmetries of each state.
The well-known d-wave superconducting order parameter is written as
\begin{equation}
\langle \Delta^{+}_{d-sc} \rangle
=-\frac{1}{\sqrt{2}} \sum_{{\bf k}} d({\bf k}) \langle c^{\dagger}_{{\bf k} \uparrow}
c^{\dagger}_{-{\bf k} \downarrow} \rangle.
\end{equation}
On the other hand, the ddw and electronic nematic order parameters are given by
\begin{eqnarray}
\langle \Delta_{ddw} \rangle &=& \frac{i}{2}\sum_{{\bf k}\sigma} d({\bf k})
\langle c^{\dagger}_{{\bf k}\sigma} c_{{\bf k}+{\bf Q}\sigma} \rangle, \nonumber\\
\langle \Delta_{nem} \rangle &=&  
\frac{1}{2}\sum_{{\bf k}\sigma} d({\bf k}) \langle c^{\dagger}_{{\bf k}\sigma} c_{{\bf k}\sigma} \rangle,
\end{eqnarray}
where $d({\bf k}) = \cos{(k_x)} - \cos{(k_y)}$, $\sigma$ represents up- and down-spin,
and ${\bf Q}=(\pi,\pi)$.
We set $a$, the lattice constant of 
a two-dimensional square lattice, to be unity.
Since the ddw order parameter is a complex value defined at the wavevector ${\bf Q}$, 
it breaks translational, time-reversal, and $\pi/2$-rotational symmetries, while the nematic state breaks
only $\pi/2$-rotational symmetry.

It was shown that there is O(4) invariance at a transition between the d-wave superconducting
and ddw states.\cite{nayak00PRB,nayak00PRB2} 
Below we show that there is SO(4) invariance at a transition between the nematic state 
and the d-wave superconductor, where the pseudospin SU(2) generators are s-wave pairing and density operators; 
the spin SU(2) and pseudospin SU(2) forms SO(4).
We discuss the difference between our finding and
the previous O(4) symmetry at a transition between the ddw and d-wave superconducting states.  
We then present a relation between the nematic and ddw states, and its competition between them.
%It has been well known that 
%the ddw ordered state is characterized by a broken time reversal,  translational
%and  $\pi/2$ rotational symmetry, while the nematic phase is characterized
%by a broken $\pi/2$ rotational symmetry of the square lattice, which are believed
%to be captured by the above order parameters.  
%
%We will first present a relation between the d-wave superconducting and electronic nematic states,
%and then a relation between the electronic nematic and ddw states.

{\it Pseudospin SU(2) generators } ---
To understand the relation between the nematic and d-wave superconducting states, 
let us first review
a similar relation found between the ddw and the d-wave superconductor, where the pseudospin
generators are $\eta$ pairing operators. 
The pseudospin $\eta$ operators was first discussed by Yang in the Hubbard model.\cite{yangPRL89}
$\eta^+$, $\eta^{-} =(\eta^+)^{\dagger}$, and $\eta_z$  are defined as follows.
\begin{eqnarray}
\eta^+ &=& \sum_{\bf k} c^{\dagger}_{{\bf k} \uparrow} c^{\dagger}_{-{\bf k}+{\bf Q} \downarrow},
\nonumber\\
\eta_z &=&  \sum_{{\bf k}} \left( c^{\dagger}_{{\bf k} \uparrow} c_{{\bf k} \uparrow}
 + c^{\dagger}_{{\bf k}+{\bf Q} \downarrow} c_{{\bf k} +{\bf Q} \downarrow} -1 \right).
\end{eqnarray}
Note that these operators form SU(2) algebra.
% ( $[\eta^+,\eta^-]= 2 \eta_z$ ).
It was shown that the $\eta$-pairing state is an eigenstate of
the Hubbard Hamiltonian. It is interesting to note that the $\eta$-pairing state
is a finite center-of-mass momentum pairing state (FFLO) of
s-wave superconductors with momentum ${\bf Q}= (\pi,\pi)$. It was later  proved
that the $\eta$-pairing state with a finite Zeeman field
can be mapped to  the Nagaoka ferromagnetic state  with a finite doping by a particle-hole
transformation, which simultaneously maps the negative-U Hubbard model to the 
positive-U Hubbard model, respectively.\cite{singhPRL91}
It was also shown that the on-site s-wave pairing operator and charge density
wave operator can be rotated into each other by the pseudospin SU(2) generators,
which is summarized by the following relation.\cite{zhangPRL90}
\begin{equation}
[ \eta^+, \rho_{\bf Q} ] = \sqrt{2} {\tilde \Delta}^+_{ssc},
\end{equation}
where $\rho_{\bf Q} =\frac{1}{2} \sum_{{\bf k} \sigma} c^{\dagger}_{{\bf k} \sigma}
c_{{\bf k}+{\bf Q} \sigma}$, ${\tilde \Delta}^+_{ssc} = 
-\frac{1}{\sqrt{2}}\sum_{\bf k} c^{\dagger}_{{\bf k}
\uparrow} c^{\dagger}_{-{\bf k} \downarrow}$, 
and ${\tilde \Delta}^{-}_{ssc} = -({\tilde \Delta}^{+}_{ssc})^{\dagger}$.

Following Yang, the pseudospin SU(2) symmetry was adapted to a critical point between the
d-wave superconductor and the ddw state.\cite{nayak00PRB}
The generators, $i \eta^{+}, i \eta^{-},
\eta_z$,  were defined to have the same forms of
$\eta$. However, there is a difference: the factor of $i$ in $\eta^{\pm}$
was introduced due to the factor $i$ in the ddw operator.
The rotation between 
%An SO(4) invariance at a transition between 
the ddw and d-wave superconducting operators can be
captured by the following commutation relation.
\begin{equation}
[ i \eta^{+}, \Delta_{ddw} ] = \sqrt{2} \Delta^{+}_{d-sc},
\label{ddw-dsc}
\end{equation}
where 
$  \Delta_{ddw}  = \frac{i}{2} \sum_{{\bf k} \sigma} d({\bf k})
 c^{\dagger}_{{\bf k}+{\bf Q} \sigma} c_{{\bf k} \sigma}$,
$  \Delta^{+}_{d-sc}  =  -\frac{1}{\sqrt{2}}\sum_{{\bf k}} d({\bf k})
c^{\dagger}_{{\bf k} \uparrow} c^{\dagger}_{-{\bf k} \downarrow}$,
and $\Delta^-_{d-sc} = - (\Delta^{+}_{d-sc})^{\dagger}$.
O(4) invariance at a transition between the ddw and d-wave superconducting state
was further discussed in Ref. \cite{nayak00PRB}.

{\it Rotation between the nematic and d-wave superconducting operators} ---
It is straightforward to find a similar relation between 
the nematic and d-wave pairing operators, where 
the pseudospin SU(2) generators are 
\begin{eqnarray}
L_+& \equiv &\Delta^{+}_{s-sc} =   \sum_{{\bf k}}
c^{\dagger}_{{\bf k} \uparrow} c^{\dagger}_{-{\bf k} \downarrow},
\nonumber\\
L_{-} & \equiv &  \Delta^-_{s-sc}= (\Delta^+_{s-sc})^{\dagger},
\nonumber\\
L_{0}& \equiv  & \Delta_z = \frac{1}{2}\sum_{{\bf k}\sigma} 
c^{\dagger}_{{\bf k} \sigma} c_{{\bf k} \sigma}-N,
\end{eqnarray}
where $N$ is the total number of lattice sites.
Note that the operators $\Delta^{+}_{d-sc}(m=1)$, $\Delta^-_{d-sc} (m=-1)$,
and $\Delta_{nem} (m=0)$ form an irreducible tensor of rank $l=1$ under the SU(2) algebra,
as follows.
\begin{eqnarray}
[ L_{\pm}, \Delta_m ] &= &  \sqrt{l(l+1)-m(m\pm 1)} \Delta_{m \pm 1},
\nonumber\\
\left[ L_{0} ,  \Delta_m \right] & =& m \Delta_m,
\label{nematic-dsc}
\end{eqnarray}
%where $\sqrt{2}=\sqrt{l(l+1)-m(m\pm 1)}$ with $l=1$ and $m=0$.
where $l=1$.
%where
%$ \Delta_{nem}  =  
%\frac{1}{2}\sum_{{\bf k},\sigma} d({\bf k}) c^{\dagger}_{{\bf k},\sigma} c_{{\bf k},\sigma}$.
Therefore, the nematic and d-wave superconducting operators can be rotated into each
other by the pseudospin generators;
\begin{equation}
[ \Delta_{s-sc}^{+}, \Delta_{nem}]= \sqrt{2} \Delta^{+}_{d-sc}.
\end{equation}

The above equation implies that there is SO(4) invariance at a transition between
the nematic and d-wave superconducting (nematic-dsc) states. 
However, a small symmetry breaking term can be present,
which will favor one state over the other. 
For example, it is possible that potential terms 
such as $-g (\Delta_{nem}^2 - \Delta_{d-sc}^2)$ can be present, 
which favors the nematic phase (the d-wave superconductor)
for $g > 0 (g < 0)$. Then, a different symmetry breaking term such as a finite chemical potential
can lead to a transition from the nematic state to the d-wave superconducting state, so
there is SO(4) symmetry at a bi-critical point.

Similar scenarios were proposed in the 
previous study of the ddw and d-wave superconducting (ddw-dsc) transition, as well as
in SO(5) theory of antiferromagnetic and d-wave superconducting states. 
However, there is a crucial difference between the ddw-dsc and nematic-dsc transitions.
In the case of the O(4) symmetry at a transition between
the ddw and d-wave superconducting state, the chemical potential is {\it a} symmetry breaking term.
Since the  chemical potential couples to one of the SU(2) generators ($\propto \mu \eta_z$),
a finite chemical potential favors the d-wave superconducting state over the ddw state.
Therefore, if an effective interaction is O(4) invariant, the system equally favors the ddw and d-wave
superconducting states at the half filling with a tight-binding dispersion.
On the other hand, if an effective interaction favors the ddw state at the half-filling, 
there is a first order transition
from the ddw state to the d-wave superconducting state at a finite chemical potential, which is
like a spin-flop transition.
Note that the nearest neighbor hopping term is O(4) invariant. 
In other words, the nearest neighbor hopping term commutes with $\eta$ pairing operator
which results from $\epsilon_{\bf k} = -\epsilon_{{\bf k}+{\bf Q}}$ where $\epsilon_{\bf k} = -2 t (\cos{k_x a}
+\cos{k_y a})$.
%A consequence of this symmetry is that the critical point occurs at exactly half-filling
%at $\mu= U/2$ in the Hubbard model where $\mu$ is the chemical potential and $U$
%is the on-site Hubbard interaction term.
%A further symmetry breaking term such as $u$ term in
%\onlinesite{Nayak} is required to tune the transition point to a finite doping concentration
%away from the half filling.

%Now let us understand back to our discussion about the consequence of the difference in pseudospin generators.

The  nearest neighbor hopping term has a quite different effect on
the nematic and d-wave superconducting transition. 
Since the pseudospin generator is an s-wave pairing operator,
the nearest neighbor hopping term is a symmetry breaking term,
in addition to the chemical potential term. 
The nearest neighbor hopping term can be written as 
\begin{equation}
H_0 = \sum_{\bf k} \epsilon_{\bf k} \psi^{\alpha \sigma \dagger}_{\bf k}  \tau_{3 \alpha}^\beta 
\psi_{\beta \sigma {\bf k}},
\end{equation}
where $\psi_{\bf k}$ the vector field with 4-components,  $\alpha(\beta)$ represents pseudospin index,
 $\tau$ is a Pauli matrix for pseudospin, and $\sigma$ denotes spin index.
%similar to Nambu representation
%in spin triplet superconductors, 
Since the electron field forms a doublet under the
pseudospin $SU(2)$ in addition to the spin $SU(2)$, $\psi_{\bf k}$ is defined as
\begin{equation}
\psi_{\bf k} = \left( c_{{\bf k} \uparrow}, c_{{\bf k} \downarrow}, c^\dagger_{-{\bf k} \downarrow},
c^\dagger_{-{\bf k} \uparrow} \right).
\end{equation}

Therefore, when an interaction equally favors the nematic and d-wave superconducting state,
the d-wave superconducting state wins over the nematic due to the presence of a
nearest neighbor kinetic term even at $\mu=0$.
An introduction of the chemical potential further favors the d-wave superconducting state, because
again it couples to the $\Delta_z$ operator.
Therefore, the realization of the nematic state in a realistic system requires  
a potential term which strongly favors the nematic state over
d-wave superconducting state in order to compensate for the effect of a nearest neighbor hopping term.
The coexistence of nematic and d-wave superconducting phases has been found by a mean field
theory in Ref. [\onlinecite{yamasePRB07}], where a strong nematic-favoring interaction was used.
This is consistent with our finding that the nearest neighbor hopping term is a pseudospin
symmetry breaking term, and a strong nematic interaction is required to compensate for its effect.

%An SO(4) invariance at a transition between the nematic and d-wave superconducting states
%can be found if one identifies a Hamiltonian which commutes with $L_0$ and $L^2$ and to show
%that $L_{\pm}$ commutes with the Hamiltonian at the critical point. 

%The same gauage transformation leads to 
%an SO(4) invariance at a transition between the nematic and d-wave superconducting states.

{\it Unitary transformation between the nematic and the ddw operators} ---
Now one may ask about a relation between two different rotations; Eq. \ref{nematic-dsc}  and  
 Eq. \ref{ddw-dsc}.
We will show below that these two equations transform from
one to the other by a unitary transformation. Therefore, if a Hamiltonian 
is invariant under such a  transformation, two states (nematic and ddw) are exactly degenerate.

We consider the following unitary transformation.
%Here we show how different operators such as $\Delta_{nem}$,
%$\Delta_{ddw}$, $\Delta_{s-sc}$, $\Delta_{d-sc}$, and $i \eta$ 
%transform from one to another under 
\begin{equation}
U^{\dagger} c^{\dagger}_{l} U = e^{i (-1)^l \frac{\pi}{4}} c^{\dagger}_l,
\label{gauge}
\end{equation}
where $l$ denotes a  lattice site.
$U= e^{i \frac{\pi}{4} (N_A - N_B)}$ where $N_A$ and $N_B$ are total number operator for 
sublattice $A$ and $B$ sites.
Under the above transformation, it is straightforward to show that the ddw and nematic
state can be smoothly rotated;
\begin{equation}
U^{\dagger} \Delta_{ddw} U =  \Delta_{nem}.
\end{equation}
What is the  significance of the above relation between the operators?
An importance of the relation is that,
if and only if a Hamiltonian of interest is invariant under the transformation, i.e.,
$U^{\dagger} H U = H$, then
\begin{eqnarray}
\langle \phi_1 | H |\phi_1  \rangle &=& \langle \phi_2 | H |\phi_2  \rangle  \nonumber\\
\langle \phi_1 |\Delta_{ddw} | \phi_1 \rangle &=& \langle \phi_2 | \Delta_{nem} | \phi_2 \rangle,
\end{eqnarray}
 where
$|\phi_1\rangle$ and $|\phi_2 \rangle $ are smoothly connected by a rotation,
$ U |\phi_1 \rangle  = |\phi_2 \rangle$, and exactly degenerate.
Therefore, the ddw and nematic states are exactly degenerate.
What breaks this degeneracy?
%Do $|\phi_1 \rangle $ and $|\phi_2 \rangle $ break different symmetries?

\begin{figure}[htb]
\epsfxsize=10cm
\includegraphics*[angle=0, width=1.0\linewidth, clip]{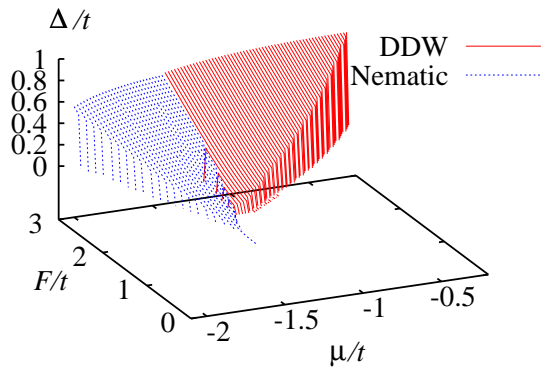}
%\epsffile{energy_mf_2.eps}
\caption{The ddw and nematic order parameters as a function of
chemical potential $\mu$ and the effective interaction $F$.
We set $t=1$, $t^{\prime} =-0.4 t$, and $t^{\prime \prime} =0$.
The red and blue regions are the ddw and nematic phases, respectively.
\label{fig:meanfield}
}
\end{figure}

Since the unitary transformation involves the sublattices of $A$ and $B$, any types of density-density or 
spin-spin interactions do not break the degeneracy. However the kinetic term does.
The most important and relevant term which breaks the degeneracy is
the nearest neighbor hopping term which is not invariant under this transformation.\cite{footnote}
Therefore, one of the states always wins over the other
due to the presence of a non-zero nearest neighbor hopping integral in realistic systems.
While the next-nearest ($t^{\prime}$) and next-next-nearest ($t^{\prime \prime}$)
hopping terms, and chemical potential are invariant under the unitary transformation,
the effect of $t$ on selecting a state can be changed in the presence of these terms.
The energetic difference between the ddw
and nematic states is presented in Fig. \ref{fig:meanfield} 
using mean-field theory for a given set of hopping parameters
and an interaction $F$ which equally favors the nematic and ddw states.
Here we set $t=1$, $t^{\prime} =-0.4 t$ and $t^{\prime \prime} = 0$,
and show how the state is stabilized
as a function of the chemical potential $\mu$ and the effective interaction $F$.

As it is shown, the ddw state is favorable at relatively low doping,
while the nematic state wins at higher doping.
The origin of the phase transition is related to a change in the density
of states, since the ddw state is favorable near doping with the nesting, $\epsilon_{\bf k}
=-\epsilon_{{\bf k}+{\bf Q}}$, while the nematic state is favorable near doping with
a van Hove singularity.   A reasonable amount of $t^\prime$ moves the nematic state
to higher doping by shifting a van Hove singularity, which is a way to avoid 
the competition with the ddw state. 
The two phases are separated by the first order phase transition; this
occurs around $\mu = - 1.4$ which corresponds to the hole doping of $0.2$,
and is almost independent of the strength of the effective interaction $F$.

{\it Summary and Discussion} ---
Finite angular momentum condensates have been studied to understand various exotic phases
in strongly correlated systems.
In particular,  condensates of particle-particle, or particle-hole pairs of $l=2$ 
relative angular momentum have been proposed in the context of high Tc cuprates. 
We studied the relations between different $l=2$ condensates; the d-wave
superconducting, ddw, and electronic nematic states proposed to be relevant
phases of underdoped cuprates.

We showed that there is SO(4) invariance at a transition  between the electronic nematic
and d-wave superconducting state. The pseudospin SU(2) generators that rotate
the nematic to d-wave pairing operators are s-wave pairing and electron density 
operators.  The important difference between a similar O(4) invariance 
at a transition between the ddw and d-wave
superconducting state transition is that the nearest neighbor hopping term is a symmetry breaking term,
which in turn always favors the d-wave superconducting state over the nematic state
even at $\mu=0$. 
A finite chemical potential further favors the d-wave superconducting state. 

We also found that the electronic nematic
operator and the ddw operator transform into each other under a unitary transformation. 
%(the $\eta$ pairing operator transforms to the s-wave pairing operator).
Therefore, if Hamiltonians are invariant under the unitary transformation, the ddw and nematic states
are exactly degenerate.  The most important and relevant term which breaks the degeneracy is
the nearest neighbor hopping integral. Since the nearest neighbor hopping is finite in 
realistic materials of our interest, one of the two states is always energetically lower than
the other.  While the chemical potential term is
invariant under the unitary transformation, the role of $t$ on its energetic selection changes
as one changes $\mu$.
We found that the ddw state is stabilized over the nematic at lower chemical potential, while
the nematic wins over the ddw state for higher chemical potential within a mean field approximation,
when the interactions equally favor these two states. 

A further study on the competition between the nematic and d-wave superconducting states may
lead us to understand a series of anisotropic scattering patterns observed in YBa$_2$Cu$_3$O$_{7-\delta}$. 
\cite{hinkov07,hinkov08,stock06,yjkao05,hamase06}
The competition between the ddw, nematic, and d-wave superconducting states beyond
the mean field theory is also an important issue, which we will
address in the near future.\cite{hdoh08}

{\it Acknowledgement} We thank Sung-Sik Lee, Arun Paramekanti and Michael Lawler for insightful discussions.
HYK thanks Aspen Center for Physics for their hospitality where this work was initiated.
This work is supported by NSERC of Canada, Canadian Institute for Advanced Research, and Canada Research Chair.


\begin{thebibliography}{100}
\bibitem{footnote} At a limit of $t \rightarrow 0$, the ddw order parameter is ill-defined,
since the ddw order parameter is defined via a current on each link between the nearest neighbor bonds.
Therefore, the ddw order parameter cannot represent translational and time-reversal symmetries,
and it is nothing but a different gauge choice of the nematic state.
\end{thebibliography}
\end{document}